\documentclass[a4paper,10pt]{article}
 
\usepackage{graphicx}
\usepackage{dcolumn}
\usepackage{bm}
\usepackage{amsmath}
\usepackage{amssymb}  

\baselinestretch
\begin{document}
\title{ A Novel Design of Dielectric Perfect Invisibility Devices}
\author{T. Ochiai$^1$, U. Leonhardt$^2$ and J.C. Nacher$^3$}   

\maketitle

\begin{center}
{\it $^1$ Faculty of Engineering, Toyama Prefectural University }
\end{center}
\begin{center}
{\it 5180 Kurokawa Imizu-shi Toyama, 939-0398, Japan}
\end{center}
\begin{center}
ochiai@pu-toyama.ac.jp
\end{center}

\begin{center}
{\it $^2$ School of Physics and Astronomy, University of St Andrews
}
\end{center}
\begin{center}
{\it North Haugh, St Andrews, KY16 9SS, Scotland}
\end{center}
\begin{center}
ulf@st-andrews.ac.uk
\end{center}

\begin{center}
{\it $^3$ Department of Complex Systems, Future University-Hakodate}
\end{center}
\begin{center}
{\it 116-2 Kamedanakano-cho Hakodate, Hokkaido, 041-8655, Japan}
\end{center}
\begin{center}
nacher@fun.ac.jp
\end{center}

\abstract{
The aim of an invisibility device is to guide light around any object put inside, 
being able to hide objects from sight. In this work, we propose a novel design of 
dielectric invisibility media based on negative refraction and optical conformal 
mapping that seems to create perfect invisibility. This design has some advantages and more relaxed 
constraints compared with already proposed schemes. In particular, it represents an example where 
the time delay in a dielectric invisibility device is zero. Furthermore, due to impedance matching 
of negatively refracting materials, the reflection should be close to zero. These findings strongly 
indicate that perfect invisibility with optically isotropic materials is possible. Finally, the area 
of the invisible space is also discussed.


}

\section{Introduction}

In Optics, Fermat's principle describes the trajectory of the light 
as the path taken between two points by light waves that can be traversed 
in the least time \cite{optics, landau}. Each dielectric medium is 
characterized 
by a refractive index $n$. This index profile integrated along the ray 
trajectory 
defines the path length. The value of the refractive index indicates a 
property 
of a material that changes the speed of light, computed as the ratio of the 
speed of 
light in a vacuum to the speed of light through the material. When light 
travels at 
an angle between two different materials, their different refractive 
indices determine 
the angle of transmission (refraction) of the ray of light. This 
relationship is known 
as Snell's law and illustrates that light changes the trajectory to 
minimize the time 
for travel between two points. A well-known example of the optical effect 
caused by 
light passing through two media with different refractive indices is a 
mirage in the desert. The 
difference in air density changes the refractive indices and makes it 
possible to see images 
from the sky above the sand. This is a consequence of the Fermat's Principle, 
because the 
light rays follow the shortest path defined by the lowest refractive 
profiles. 

Our invisibility problem has strong links with these principles 
of Optics. Recently, a few models have been proposed to create 
a perfect illusion of invisibility. The main idea is to design 
a dielectric medium composed of a specific refractive index that literally 
bends and guides 
the light around a desired object. As the device itself should be 
invisible, 
an external observer would not see the object. The optical effect will be 
equivalent 
to observe the light rays propagating across empty space, because objects 
in the background 
will be visible. 
Until now, a perfect invisibility effect based 
on isotropic media has not been achieved and the ultimate wave nature of 
light has been argued as the main reason.
This issue has been extensively debated
in several works \cite{nach,wolf,leon1,leon2}. However, 
using anisotropic media these distortions could be reduced to zero, in principle \cite{pendry2006}. Furthermore, a recent experimental demonstration
of a cloaking device based on artificially structured metamaterials \cite{exp1} has succeed 
in decreasing scattered waves emitted from the hidden object. 
The object remains invisible over a narrow band
of microwave frequencies. More recently, a new design of a non-magnetic optical cloak 
that reduces the scattered waves, and extends the operative range of frequencies to the optical spectrum, 
was proposed and investigated using computer simulations 
\cite{exp2}. Implications of these research lines, future trends and remaining problems on the field are commented and 
summarized in \cite{exp3}.

The discovery of new materials (also known as {\it metamaterials}) 
is opening up exciting avenues for new electromagnetic applications in 
diverse areas, such as medical imaging, microwaves, lenses, 
radars, defense and telecommunications. These new features cannot be 
derived using materials found in nature. Metamaterials 
then represent artificial media with unusual electromagnetic
properties. These material properties emerge as a consequence of their 
complex and often periodic structures rather 
than their chemical composition. Therefore, engineered materials with 
pre-designed periodic nanostructures 
could lead to a large range of refractive indices with a rich and great 
variety of new electromagnetic properties. This 
is also true for the recently discovered metamaterials with negative 
refractive index \cite{meta}. A striking 
property of materials with negative index is that the rays will be 
refracted on the same side 
of the normal on entering the material according to the Snell's law. The connections between
invisibility and negative refraction were described in \cite{exp4} using the formalism of general relativity. 

In this work, we extend the original idea of one of us \cite{leon1,leon2} 
based on optimal conformal 
mapping to create dielectric invisibility
devices.
 In this context, we propose a new design for a cloaking device 
made of a metamaterial 
with negative refractive index. 
Our scheme reduces the time delay caused by the device to zero
and relaxes the constraints imposed by the requirement for bounded orbits in 
the Riemann sheet.
By following \cite{leon1,leon2}, a
dielectric medium conformally maps a physical space $z$ onto Riemann sheets 
given by an analytic function $w(z)$. As 
a feature of conformal maps, the angles between coordinate lines are 
conserved. By using, for example, the 
well-known Joukowski transformation \cite{book1,book2}, 
the papers \cite{leon1,leon2} exploit a 
mapping 
between the physical space described by the complex field $z$ and the 
analytic function $w(z)$ 
that represents the mathematical space composed of two Riemann sheets and 
a branch cut which connects both sheets. As a result 
any object located at the center of physical space cannot be detected by an 
external observer. In other words, 
it seems hidden inside the Riemann sheets. 

The most important distortions of invisibility are caused by reflections 
and time delays \cite{leon2}. In this work, 
we investigate how to relax the constraints for designing and manufacturing 
invisibility devices. In previous works, 
refractive-index profiles on the interior Riemann sheet should guide the 
rays around the branch points 
and should define close loops or trajectories. Therefore, according to 
classical dynamics \cite{landau}, the number 
of potential or index profiles is drastically reduced to a few ones: 
harmonic-oscillator, Kepler profile and Maxwell's fish eye. In contrast, 
our proposed design is able to generate a large variety of bounded orbits 
for rays using striking properties 
of negative refractive indices. More importantly, the proposed device based on negatively refracting material 
represents an example where the time delay in a dielectric invisibility device is zero. Furthermore, 
due to impedance matching of negatively refracting materials the reflection should be close to zero. These 
findings strongly indicate that perfect invisibility with optically isotropic materials is possible. Finally, 
the area of the invisible space is also discussed.



\section{Mathematical Formulation}

In this section, we follow refs. \cite{leon1,leon2} and 
describe 
the rays of light based on the Hamilton's analogy \cite{optics} between the 
trajectory 
of rays in media and the motion of particles governed by classical 
mechanics principles. 
Let us assume that the refractive index $n(\mathbf{x})$ does not vary too 
much comparing 
with the wavelength of light.

Then, the electromagnetic wave in media can be expressed using the Helmholtz equation:

\begin{eqnarray}\label{eqn: helmholtz equation} 
( \Delta+ \frac{n^2\omega^2}{c^2})\psi=0.
\end{eqnarray}

Next, we introduce the effective time $\tau$ measured in spacial units
\begin{eqnarray}
d\tau=\frac{c}{n^2}dt,
\end{eqnarray}
it is possible to derive the trajectory of rays as follows 
\cite{landau,leon1}:
\begin{eqnarray} \label{eqn: newton equation}  
\frac{d^2\mathbf{x}}{d\tau^2}=\frac{1}{2}\triangledown n^2.
\end{eqnarray}
This equation resembles the Newton's second law equation for light rays. 
Integrated (\ref{eqn: newton equation}) by $\mathbf{x}$, we obtain 
\begin{eqnarray}\label{eqn: enegry equation}
\frac{1}{2}(\frac{d\mathbf{x}}{d\tau})^2-\frac{n^2}{2}=\mbox{constant}.
\end{eqnarray}
The corresponding energy $E$ and potential $U$ follows the relationship
 \cite{optics,leon1}:
\begin{eqnarray}
U-E=-\frac{n^2}{2}.
\end{eqnarray}
Here, we can show that this constant in the right hand side of (\ref{eqn: 
enegry equation}) vanishes as follows:
\begin{eqnarray}\label{eqn: enegry equation2}
\frac{1}{2}(\frac{d\mathbf{x}}{d\tau})^2-\frac{n^2}{2}=\frac{1}{2}(\frac{dx}{d\tau})^2-(E-U)=0.
\end{eqnarray}
Therefore, from (\ref{eqn: enegry equation2}) we obtain
\begin{eqnarray}
|\frac{d\mathbf{x}}{d\tau}|=n .
\end{eqnarray}
In what follows, we use the Newton's second law equation for light rays 
(\ref{eqn: newton equation}) 
for designing invisible devices.

\section{Conformal mapping}
In this section, we review the conformal mapping introduced by 
\cite{leon1,leon2} in the context of invisibility.
This mapping is used to design and create the invisible space inside the 
material. The property of this conformal 
mapping and the associated Riemann surfaces plays a crucial role to design 
the cloaking devices.

\subsection{Complex plane}
We assume that the medium is uniform along the $z$-direction. Then, we 
introduce 
the complex coordinate $z=x+iy$ and $\bar{z}=x-iy$. The derivative of 
complex coordinate is given by 
\begin{eqnarray}
&&\frac{\partial}{\partial z}=\frac{1}{2}(\frac{\partial}{\partial 
x}+\frac{1}{i}\frac{\partial}{\partial y}),\\
&&\frac{\partial}{\partial\bar{z}}=\frac{1}{2}(\frac{\partial}{\partial 
x}-\frac{1}{i}\frac{\partial}{\partial y}).
\end{eqnarray}
Next, using complex coordinates, Helmholtz equation (\ref{eqn: helmholtz 
equation}) is transformed into
\begin{eqnarray}\label{eqn: helmholtz equation complex}
(4\frac{\partial^2}{\partial 
z\partial\bar{z}}+\frac{n^2\omega^2}{c^2})\psi=0.
\end{eqnarray}

Let us now consider the analytical function $w=w(z)$, which is conformal 
mapping. Then, by using
\begin{eqnarray}
&&\frac{\partial}{\partial z}=\frac{\partial w}{\partial 
z}\frac{\partial}{\partial w}\\
&&\frac{\partial}{\partial\bar{z}}=\frac{\partial \bar{w}}{\partial 
\bar{z}}\frac{\partial}{\partial \bar{w}},
\end{eqnarray}
we find that, by  the conformal mapping $w=w(z)$, the Laplacian operator is 
transformed as follows
\begin{eqnarray}
\frac{\partial^2}{\partial z\partial\bar{z}}=|\frac{\partial w}{\partial 
z}|^2\frac{\partial^2}{\partial z\partial\bar{z}}.
\end{eqnarray}

When the refractive index is transformed as 
\begin{eqnarray}\label{eqn: transformation of n}
n_z=n_w|\frac{dw}{dz}|
\end{eqnarray}
by conformal mapping, the Helmholtz equation (\ref{eqn: helmholtz equation 
complex}) is invariant. 
Here, $n_z$ and $n_w$ are the refractive index of $z$ and $w$ complex 
plane, respectively. In 
what follows, we use this invariance to design the invisible device.

\subsection{Conformal mapping and Riemann surface}
We map the physical space $z$ into Riemann surface $w$ by using the 
following conformal map:
\begin{eqnarray}\label{eqn: conformal map}
w=z+\frac{a^2}{z}.
\end{eqnarray}
The inverse function of (\ref{eqn: conformal map}) is given by
\begin{eqnarray}
z=\frac{1}{2}(w\pm\sqrt{w^2-4a^2}).
\end{eqnarray}
Here, $w$-space is a Riemann surface which consists of two sheets. By the 
conformal map (\ref{eqn: conformal map}), the 
outside of the circle $|z|>a$ is mapped into the first Riemann sheet in 
$w$-space, and the inside 
of the circle $|z|<a$ is mapped into the second Riemann sheet in $w$-space.
 The branch cut is given by $-2a< \mathrm{Re}(w) <2a, 
 \mathrm{Im}(w)=0$ in $w$-space, which 
corresponds to the circle $|z|=a$ in $z$-space. 

\subsection{Property of conformal mapping}
When we take $|z|\to \infty$, the conformal map (\ref{eqn: conformal map}) 
is simplified as follows:  
\begin{eqnarray}
w\sim z.
\end{eqnarray}
Therefore, in the asymptotical region $|z|\to \infty$, $z$-space and the 
first Riemann sheet of $w$-space 
can be identified with each other. 

We set $z$ as polar cordinate $z=r e^{i\theta}$ and $w$ as cartesian 
coordinates $w=u+iv$, then 
the conformal map (\ref{eqn: conformal map}) maps the circle ($r=r_0$) in 
physical $z$-space into the ellipse:
\begin{eqnarray}\label{eqn: elliptic curve}
\frac{u^2}{(a^2/r_0+r_0)^2}+\frac{v^2}{(a^2/r_0-r_0)^2}=1
\end{eqnarray}
in $w$-space. This property will be used later.

\section{New model of cloaking devices}

\subsection{Negative-refraction metamaterials}
Metamaterials are man-made media where electromagnetic waves do not behave 
as usually expected. As a consequence, new 
electromagnetic properties and effects may emerge and be observed. This 
term is particularly used when the material 
has properties not found in naturally-formed substances. Therefore, 
engineered materials 
with complex and pre-designed nanostructures could lead to a large 
range of refractive indices with 
a rich and great variety of new electromagnetic properties in many diverse 
fields as microelectronics, medical imaging 
as well as defense and telecommunications \cite{pendry2006, 
meta}. This is 
also true for the recently discovered metamaterials with negative 
refractive index. A striking 
property of materials with negative index is that the rays will be 
refracted on the same side 
of the normal on entering the material according to the Snell's law. Then, 
it allows to use this 
property to design novel cloaking devices. Here, we propose a device with 
two Riemann sheets. While 
the first Riemann sheet the material has refractive index $n$=1, the second 
Riemann sheet is composed of
four quadrants with alternative values negative and positive refractive 
index profile $n$. As a consequence, 
the existence of quadrants with different refractive index makes it 
possible to generate bounded orbits 
with more relaxing constrains than suggested previously \cite{leon1,leon2}. 
By using the 
negative refractive index, the trajectory around branch cut points in the 
second riemann sheet 
can be seen as a mirror-like reflection of light. We will explain this in 
detail in the next section.

\subsection{Refractive index}
We first define the refractive index $n_w$ on Riemann suface $w$ 
defined by the conformal mapping (\ref{eqn: conformal map}). We can obtain 
the refractive index $n_z$ on physical  
space $z$ by using transformation shown in (\ref{eqn: transformation of 
n}). In 
what follows, for the matter of convenience, we write just $n$ for both 
refractive index $n_z$ and $n_w$, 
when the meaning is clear from the context.

 The space $w$ is composed by two Riemann sheets. The refractive index of 
the first Riemann sheet in space $w$ 
 is given by 
\begin{eqnarray}
n_w=1.
\end{eqnarray}
Next, we divided the second Riemann sheet in space $w$ into four regions as 
follows:
\begin{eqnarray}
&&\mbox{Region 1 (R1):} ~~(x<2a~~ \mbox{and}~~ y>0),\\
&&\mbox{Region 2 (R2):} ~~(x>2a~~ \mbox{and}~~ y>0), \\
&&\mbox{Region 3 (R3):} ~~(x>2a~~ \mbox{and}~~ y<0),\\
&&\mbox{Region 4 (R4):} ~~(x<2a~~ \mbox{and}~~ y<0).
\end{eqnarray}
The refractive index on the second Riemann sheet in space $w$  is given by
\begin{eqnarray}\label{eqn: refractive index}
n_w&=&\sqrt{2E-2U(x)}\nonumber\\
&=&\left \{ 
\begin{array}{l}
\sqrt{2E-2A(2a-x)} ~~~~(\mbox{R1}: x<2a~~ \mbox{and}~~ y>0) \\
-\sqrt{2E+2A(2a-x)} ~~~~(\mbox{R2}: x>2a~~ \mbox{and}~~ y>0) \\
\sqrt{2E+2A(2a-x)} ~~~~(\mbox{R3}: x>2a~~ \mbox{and}~~ y<0)\\
-\sqrt{2E-2A(2a-x)} ~~~~(\mbox{R4}: x<2a~~ \mbox{and}~~ y<0)
\end{array}
\right.
\end{eqnarray}
These equations implicitly define the potential $U(x)$ of the Newton 
equation for the light ray.

It is worth noticing that the refractive index on the second Riemann sheet in space $w$ is completely antisymmetric with respect
to the line $x=2a$ and $x$-axis.


\subsection{Energy constraints}
Since the value inside the square root in the right hand side of (\ref{eqn: 
refractive index}) in $R1$ region 
should be positive ($n^2=2E-2U$ should be positive),  we obtain the 
constraint
\begin{eqnarray}\label{eqn: enegy constraint}
2E>2A(2a-x).
\end{eqnarray}
In order to satisfy the previous condition (\ref{eqn: enegy constraint}) 
for any starting position 
on branch cut $-2a<x<2a, y=0$, we have the constraint
\begin{eqnarray}\label{eqn: enegy constraint2}
E>4Aa.
\end{eqnarray}
Therefore, from (\ref{eqn: enegy constraint2}), $A$ is ranged between
\begin{eqnarray}\label{eqn: enegy constraint3}
0<A<\frac{E}{4a}.
\end{eqnarray}
In what follows, we fix the value $E$, and consider $A$ as a free parameter 
restricted by (\ref{eqn: enegy constraint3}).

\subsection{Trajectory of light rays}
We will compute the trajectory in $R1$ region. The other three regions can 
be obtained 
by taking a mirror-like reflection of the $R1$ region. This is the main 
idea of our approach. 

Next, we solve the Newton equation (\ref{eqn: newton equation}) under $n_w$ 
given in (\ref{eqn: refractive index}). For the $R1$ 
area, the trajectory is given by
\begin{eqnarray}
x&=&\frac{1}{2}A\tau^2 + v_x \tau +x_0 \label{eqn: trajectory1},\\
y&=&v_y\tau. \label{eqn: trajectory2}
\end{eqnarray}
Here, $v_x$ and $v_y$ are the initial velocity at the branch cut in second 
Riemann sheet $w$ and 
the absolute value of velocity is given by 
\begin{eqnarray}
v_x^2+v_y^2=n^2=2E-2A(2a-x_0).
\end{eqnarray}

\begin{figure}[htb]
\setlength{\unitlength}{1cm}
\begin{picture}(15,12)(-1,-1)
\put(-1,0){\includegraphics[scale=0.5]{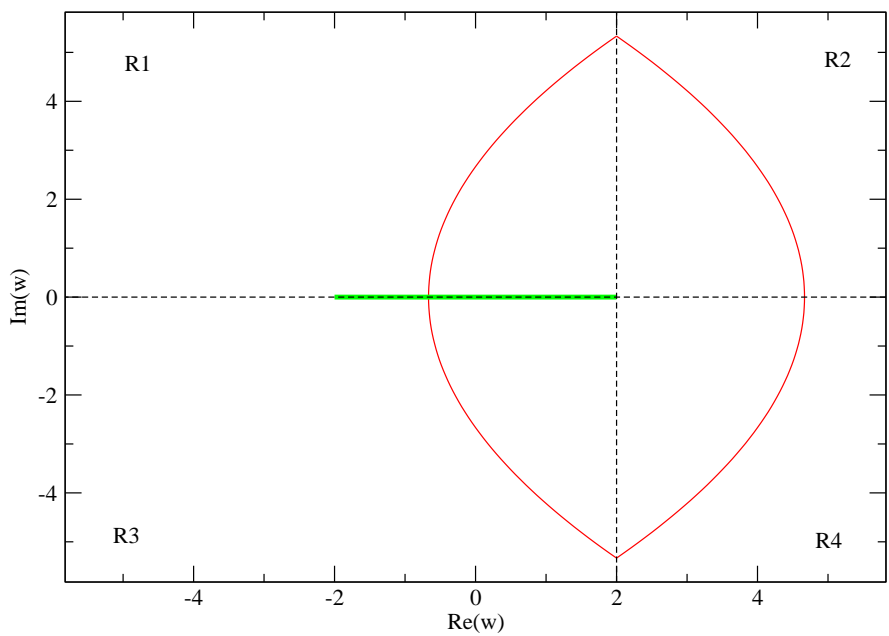}}
\end{picture} 
\caption{\small{Trajectory of light rays on the second Riemann sheet in 
space $w$. The branch is represented by a bold
line between -2 to 2 in the real axis. The four areas describing 
the refractive index profile of the cloaking device are shown in the
figure. Parameters are set to $a=1$, $E=4$.}}
\label{fig: trajectory}
\end{figure}

The novelty is that, using the negative refractive index, we can enclose 
the trajectory in the second Riemann sheet in space $w$ 
(see Fig. \ref{fig: trajectory}). Therefore, we can make the starting point 
and ending point be the same. Moreover, 
the direction of initial vector and final vector can coincide with each 
other.

In contrast, \cite{leon1,leon2} proposed a few examples to enclose the 
trajectory using bounded orbits according 
to classical mechanics: Kepler and harmonic-oscillator potentials, and Maxwell's fish eye. We think 
that our approach is more flexible 
and admits a large variety of orbits by using the negative index profile 
features. Thus, it is not limited 
to the small number of bounded orbits described by classical mechanics.

\subsection{The maximum point in the second Riemann sheet}
Let $\tau_1$ be the parameter when the light arrives to the boundary 
between $R1$ and $R2$ area. This can 
be obtained by solving
\begin{eqnarray}\label{eqn: parameter}
\frac{1}{2}A\tau_1^2 + v_x \tau_1 +x_0=2a.
\end{eqnarray}
The solution of this equation (\ref{eqn: parameter}) is given by 
\begin{eqnarray}\label{eqn:tau} 
\tau_1=\frac{-v_x+\sqrt{v_x^2+2A(2a-x_0)}}{A}.
\end{eqnarray}
From (\ref{eqn: trajectory1}), we obtain the minimum $x$-point of the 
parabolic curve in the $R1$ region as follows
\begin{eqnarray}
x_{min}=-\frac{v_x^2}{2A}+x_0.
\end{eqnarray}
From (\ref{eqn: trajectory2}), the maximum $y$-point in $R1$ region is 
given by
\begin{eqnarray}\label{eqn: ymax init}
y_{max}=v_y\tau_1=v_y\frac{-v_x+\sqrt{v_x^2+2A(2a-x_0)}}{A}.
\end{eqnarray}
In what follows, we consider the case that the light enters vertically 
$v_x=0$ for the matter of convenience. Then, (\ref{eqn: ymax init}) becomes
\begin{eqnarray}\label{eqn: ymax init2}
y_{max}=2\sqrt{(\frac{E}{A}-(2a-x_0))(2a-x_0)}.
\end{eqnarray}

We will maximize $y_{max}$, by changing the starting point $x_0$. This can 
be done 
by consider the following two cases (i) and (ii), according to $A$ value.

\noindent (i) In case of $(\frac{E}{8a}<A<\frac{E}{4a})$\\
In case of $(\frac{E}{8a}<A<\frac{E}{4a})$, (\ref{eqn: ymax init2}) takes 
the maximum  
\begin{eqnarray}\label{eqn: ymax1}
y_{max}=
\frac{E}{A}~~~~~~~~(\mbox{for}~~~~ \frac{E}{8a}<A<\frac{E}{4a})
\end{eqnarray}
when the light ray starts at $x_0=2a-E/(2a)$.

\noindent (ii) In case of $(0<A<\frac{E}{8a}$)\\
In case of $(0<A<\frac{E}{8a}$), (\ref{eqn: ymax init2}) takes the maximum  
\begin{eqnarray}\label{eqn: ymax2}
y_{max}=4\sqrt{\frac{E}{A}a-4a^2}~~~~ (\mbox{for}~~~~ 0<A<\frac{E}{8a})
\end{eqnarray}
when the light ray starts at $x_0=-2a$. This starting point is the left 
hand end point of the branch cut. 

In both cases (i) and (ii), with increasing $A$, $y_{max}$ in space $w$ 
decreases, and the invisible space of $z$ becomes larger. We will discuss this issue in more detail in section \ref{section: The size of the invisible space}.

\section{Time delay}
Due to the opposite signs in the refraction index described in the second Riemann sheet, the proposed media presents by construction an example where the time delay in a dielectric invisibility device is zero. In particular, it applies to devices based on optically isotropic materials.

The time delay is conformal invariant and can be computed by
\begin{eqnarray}
cT
&=&\int n_z |dz|\nonumber\\
&=&\int n_w |dw|.
\end{eqnarray}

Here, we compute the time delay. First, we focus on the $R1$ region. 
Then, the time delay for $R1$ region is given by 
\begin{eqnarray}\label{eqn:time delay1}
cT_{R1}
&=&\int_{C_1} n_w |dw|\nonumber\\
&=&\int_0^{\tau_1} n_w^2 d\tau \nonumber\\
&=&\frac{1}{3}A^2\tau_1^3+Av_x\tau_1^2+(2E-2A(2a-x_0))\tau_1,
\end{eqnarray}
where $C_1$ is the trajectory of the light in R1 region.
Inserting (\ref{eqn:tau}) into this equation (\ref{eqn:time delay1}), we obtain  an explicit form as follows
\begin{eqnarray}\label{eqn:time delay2}
cT_{R1}=(2E-\frac{2}{3}v_x^2-\frac{4}{3}A(2a-x_0))\frac{-v_x+\sqrt{v_x^2+2A(2a-x_0)}}{A}+\frac{2}{3}v_x(2a-x).
\end{eqnarray}

We consider the other regions. The time delay on R2 region is the same absolute value as on R1 region but with 
opposite sign, because the refractive index on R2 has the opposite sign of that on R1,
\begin{eqnarray}
cT_{R1}=\int_{C_1} n_w |dw|=-\int_{C_2} n_w |dw|=-cT_{R2}
\end{eqnarray}
where $C_2$ is the trajectory of the light in R2 region.

Similary, for R3 and R4 region, we obtain,
\begin{eqnarray}
cT_{R3}=\int_{C_3} n_w |dw|=-\int_{C_4} n_w |dw|=-cT_{R4}
\end{eqnarray}
where $C_3$ and $C_3$ are the trajectories of the light in R3 and R4 region respectively.

Therefore the total of them is completely cancelled out, and the total time delay is exactly zero.
\begin{eqnarray}
cT_{total}&=&cT_{R1}+cT_{R2}+cT_{R3}+cT_{R4}\nonumber\\
&=&0
\end{eqnarray}

\section{Reflection }

Together with time delay, reflection waves are also a distortion that prevents to achieve a perfect invisibility device. However, 
in the proposed device, due to impedance matching of negatively refracting materials, the reflection should be close zero. This 
finding, together with the zero time delay shown in the above section, strongly indicates that 
perfect invisibility is possible in isotropic media. 

New materials called metamaterials have recently been created. They are characterized by a negative refractive index. These materials 
can provide a total refraction phenomena when the wave impedances of the two media are matched. As a consequence, there 
is no reflected wave, and the distortions due to reflection are zero. Our proposed invisibility device consists of
a plane with positive/negative $n$ boundary. In this case, the reflection would be close to zero by impendance matching. To be precise, given 
two materials with a defined border, impedance matching means that $Z_{in}$=$Z_{out}$ i.e., output impedance 
of a source of waves is equal to the input impedance of the waves, and there is no reflected wave \cite{book3}.

Furthermore, it is worth noticing that our case does not contradict Nachman's theorem
about the impossibility of perfect invisibility in standard isotropic materials,
because these 
negatively refracting materials exploit the polarized nature of electromagnetic waves. It means that these 
materials are described in terms of the permittivity $\epsilon$ and the permeability 
$\mu$.

Let us illustrate the above described reflectionless property of our proposed device 
using an incident plane wave solution of Maxwell's equation propagating through a 
planar dielectric interface.

Let us first consider the reflection coefficients that can be obtained from Maxwell eqs. and the boundary conditions at dielectric interfaces (see \cite{book3}
for details). In case that $E$ is vertical to the incoming plane the coefficient reads as  
\begin{eqnarray}
\frac{E_r}{E_i}=\frac{\sqrt{\frac{\mu^\prime}{\epsilon^\prime}}\cos \phi-\sqrt{\frac{\mu}{\epsilon}}\sqrt{1-(\frac{n}{n^\prime})^2\sin^2 \phi}}{\sqrt{\frac{\mu^\prime}{\epsilon^\prime}}\cos \phi+\sqrt{\frac{\mu}{\epsilon}}\sqrt{1-(\frac{n}{n^\prime})^2\sin^2 \phi}},
\end{eqnarray}
where $\phi$ denotes the angle of incidence.
In case that $E$ is parallel to the incoming plane the coefficient is given by  
\begin{eqnarray}
\frac{E_r}{E_i}=\frac{\sqrt{\frac{\mu}{\epsilon}}\cos \phi-\sqrt{\frac{\mu^\prime}{\epsilon^\prime}}\sqrt{1-(\frac{n}{n^\prime})^2\sin^2 \phi}}{\sqrt{\frac{\mu}{\epsilon}}\cos \phi+\sqrt{\frac{\mu^\prime}{\epsilon^\prime}}\sqrt{1-(\frac{n}{n^\prime})^2\sin^2 \phi}}.
\end{eqnarray}

For a general angle of $E$, one should consider linear combination of these two cases. There are three limiting cases (1), (2) and (3):

(1) $\epsilon=-\epsilon^\prime$ and $\mu=-\mu^\prime$ case.
In this case, there is negative refraction and no reflection at all for a general angle of incidence and polarization:
\begin{eqnarray}
\frac{E_r}{E_i}=0.
\end{eqnarray}
We remark that this holds for any polarization, by considering the 
linear combination of the two cases (i) $E$ is parallel to incoming plane and (ii) $E$ is vertical to incoming plane.
 
(2) Normal incidence case. In this case, the coefficient reads as  
\begin{eqnarray}
\frac{E_r}{E_i}=\frac{\sqrt{\frac{\mu}{\epsilon}}-\sqrt{\frac{\mu^\prime}{\epsilon^\prime}}}{\sqrt{\frac{\mu}{\epsilon}}+\sqrt{\frac{\mu^\prime}{\epsilon^\prime}}}=\frac{Z_2-Z_1}{Z_2+Z_1} .
\end{eqnarray}

(3) Impedance-matched case. When $Z=Z^\prime$, the coefficient reads as  
\begin{eqnarray}
\frac{E_r}{E_i}=\frac{\cos \phi-\sqrt{1-(\frac{n}{n^\prime})^2\sin^2 \phi}}{\cos\phi+\sqrt{1-(\frac{n}{n^\prime})^2\sin^2 \phi}}.
\end{eqnarray}

In our problem, at the boundary of R1 and R2, we can take the following form, i.e., case(1):
\begin{eqnarray}
&&\epsilon_{R1}=-\epsilon_{R2}~~~~\mbox{(at the boundary between R1 and R2)}\\
&&\mu_{R1}=-\mu_{R2}~~~~\mbox{(at the boundary between R1 and R2)}
\end{eqnarray}
where $\epsilon_{R1}$ and $\epsilon_{R2}$ are the permittivity of R1 and R2 region respectively, 
$\mu_{R1}$ and $\mu_{R2}$ are the permeability of R1 and R2 region respectively.
This gives the property
\begin{eqnarray}
n_{R1}=-n_{R2}~~~~\mbox{(at the boundary between R1 and R2)}
\end{eqnarray}
where $n_{R1}$ and $n_{R2}$ are the refractive indexes of R1 and R2 regions, respectively. This reproduces 
the property (\ref{eqn: refractive index})  of our refractive index at the boundary.

In this case, there is no reflection at all for a general angle of incidence and polarization 
between the boundary of R1 and R2. In the same way, we can show that there is no reflection between the boundary of R2 and R3 (R3 and R4). 

As for the branch cut,  when vertical entering ($\phi=0$) in the case of impedance
matching ($Z_1=Z_2$), there is no reflection at the branch cut, see case
(2). Otherwise, there is some reflection, because of the discontinuous 
behavior of refractive index $n$ and $n^\prime$, see case (3). However, in order
to achieve perfect invisibility, we can solve this problem by modifying our
potential $U(x)$ as follows. Due to the striking property of negative refracting 
index, the only requirement for potential $U(x)$ in $R1$ region of the second 
Riemann sheet is that the light should be bent to the left to the right (i.e., the light 
entering from the first Riemann sheet to R1 region of the second Rieman sheet
across the branch cut ($-2a<x<2a, y=0$) should go through the boundary between R1
and R2 sheet ($x=2a, y>0$)).
Our potential is the simple example satisfying this condition. 
However, it is easy to see that there are many other suitable potentials as well. Therefore, we could slightly modify our potential around the branch cut in R1 region to smoothly
connecting the potential of the first Rieman sheet to the potential of R1
region, keeping the condition that the light should be bent from the left to
the right. Then, the refractive index can be smoothly connected. As a result, 
there is vanishing reflection around the branch cut at all.

We would like to remark two points. The first remark is that this is
possible because our proposed design is flexible enough to potentially
generate a large variety of bounded orbits for light rays using the
properties of negative refracting indices. The second remark is that this
modification does not changes the absence of time delay in our device. In other
words, even if we modify the potential, the time delay is still zero.

To summarize, we can conclude that perfect invisibility devices in isotropic media seem possible.

\section{The size of the invisible space}\label{section: The size of the invisible space}
In this section, we investigate the maximum size of invisible space. An 
object within the size of this space would be invisible and
hidden from sight.

\subsection{The radius of invisible space}
Although several local invisible areas may exist, depending on the ray light 
trajectories, we focus on identifying
the symmetric size of invisible space characterized by radius $R$.

From (\ref{eqn: elliptic curve}), the condition that the $r<R$ in physical 
space $z$ is invisible is as follows. For 
all point $(u,v)$ of the ray light trajectory in $w$-space, the following 
equation 
\begin{eqnarray} \label{eqn: invisible condition}
\frac{u^2}{(a^2/R+R)^2}+\frac{v^2}{(a^2/R-R)^2}<1
\end{eqnarray}
holds. Therefore, the radius of the invisible space is given by the minimum 
of 
\begin{eqnarray}\label{eqn: invisible condition2}
r&=&\frac{1}{2\sqrt{2}}\{ 
\sqrt{u^2+v^2+4a^2+\sqrt{(u^2+v^2+4a^2)^2-16a^2u^2}}\nonumber\\
&&-\sqrt{u^2+v^2-4a^2+\sqrt{(u^2+v^2+4a^2)^2-16a^2u^2}}
\}
\end{eqnarray} 
for all points $(u,v)$ of the ray light trajectory

The possible points for minimizing (\ref{eqn: invisible condition2}) are 
$(u,v)=(2a,y_{max})$ or $(0,6a)$, 
where $y_{max}$ is given by (\ref{eqn: ymax1}) and (\ref{eqn: ymax2}).\\
We will compute these points by considering the following three cases (i), 
(ii) and (iii), respectively.\\

\noindent (i) In case of $(u,v)=(2a,y_{max})$ and $(0<A<\frac{E}{8a})$.\\
By substituting $u=2a$ and $v=y_{max}$ into (\ref{eqn: invisible 
condition2}), we transform (\ref{eqn: invisible condition2}) into
\begin{eqnarray}\label{eqn: invisible condition3}
r_1(A)=\frac{y_{max}}{4}\{1+\sqrt{1+\frac{16a^2}{y_{max}^2}}-\sqrt{2+2\sqrt{1+\frac{16a^2}{y_{max}^2}}}\}.
\end{eqnarray}
Then, substituting (\ref{eqn: ymax2}) into (\ref{eqn: invisible 
condition3}), we obtain
\begin{eqnarray}\label{eqn:r1}
r_1(A)=\sqrt{\frac{E}{A}a-4a^2}\{1+\sqrt{1+\frac{aA}{E-4aA}}-\sqrt{2+2\sqrt{1+\frac{aA}{E-4aA}}}\}.
\end{eqnarray}
In particular, when $A\to 0$
\begin{eqnarray}
r_1(A)\sim \frac{\sqrt{A}}{4\sqrt{E}}a^{\frac{3}{2}}.
\end{eqnarray}
Thus, when $A$ is small, the radius of invisible space goes to 0.\\

\noindent (ii)In case of $(u,v)=(2a,y_{max})$ and 
$(\frac{E}{8a}<A<\frac{E}{4a})$.\\
Substituting  (\ref{eqn: ymax1}) into (\ref{eqn: invisible condition3}), we 
obtain
\begin{eqnarray}\label{eqn:r2}
r_2(A)=\frac{E}{4A}\{1+\sqrt{1+\frac{16a^2A^2}{E^2}}-\sqrt{2+2\sqrt{1+\frac{16a^2A^2}{E^2}}}\}.
\end{eqnarray}
In particular, this function takes the maximum 
\begin{eqnarray}
r_2(\frac{E}{4a})=(1+\sqrt{2}-\sqrt{2+2\sqrt{2}})a\sim 0.21a,
\end{eqnarray}
when $A=\frac{E}{4a}$.

As an example, we write the value of $r_2$ for different value of $A$. 
When $A=\frac{E}{8a}$,
\begin{eqnarray}
r_2(\frac{E}{8a})=(2+\sqrt{5}-2\sqrt{2+\sqrt{5}})a\sim 0.12a
\end{eqnarray}
which is smaller than Eq. (\ref{eqn:r3}).\\

\noindent (iii) In case of $(u,v)=(6a,0)$\\
By substituting $(u,v)=(6a,0)$ into (\ref{eqn: invisible condition2}), we 
obtain
\begin{eqnarray}\label{eqn:r3}
r_3=(3-2\sqrt{2})=0.17a.
\end{eqnarray}

\begin{figure}[htb]
\setlength{\unitlength}{1cm}
\begin{picture}(15,12)(-1,-1)
\put(-1,0){\includegraphics[scale=0.5]{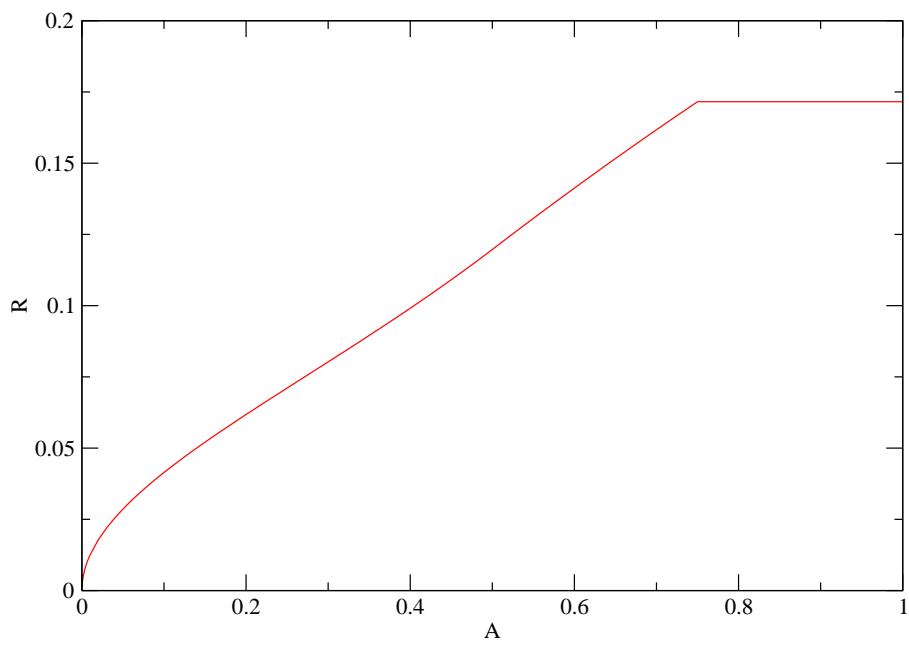}}
\end{picture} 
\caption{\small{Maximum radius of invisible space. With increasing the 
potential parameter $A$, the radius
of invisible space becomes large up to the value $0.17a$. Parameters are 
set to $a=1$, $E=4$.}}
\label{fig: radius}
\end{figure}

By summarizing (i), (ii) and (iii), we obtain the radius of the invisible 
space  $R$ as follows:
\begin{eqnarray}\label{eqn: maximum radius of invisible space}
R(A) 
&=&\left \{ 
\begin{array}{l}
r_1(A) ~~~~(\mbox{for}~~~ 0<A<\frac{E}{8a}))  \\
r_2(A)  ~~~~(\mbox{for}~~~ \frac{E}{4a}<A<\frac{3E}{16a}))  \\
r_3 ~~~~(\mbox{for}~~~ \frac{3E}{16a}<A<\frac{E}{4a})), \\
\end{array}
\right.
\end{eqnarray}
where $r_1(A)$, $r_2(A)$ and $r_3$ are given by (\ref{eqn:r1}), 
(\ref{eqn:r2}) and (\ref{eqn:r3}).
Here, we use the fact that $(r_2(\frac{3E}{16a})=r_3))$. We show the radius 
of the invisible space (\ref{eqn: maximum radius of invisible space}) in 
Fig. \ref{fig: radius}. This result 
indicates that with increasing $A$, the radius of invisible space increases 
up to $0.17a$. As the size of the invisible space is proportional 
to the branch cut $a$, the proposed invisibility device may potentially allow us 
to hide extensive objects.

\section{Summary}

In this work, we have proposed a new design for a cloaking device made of a 
metamaterial 
with negative refractive index. Our results indicate that by using negative 
refractive materials, the constraints 
for designing and manufacturing invisibility devices can be more relaxed. 
In previous works, 
refractive-index profiles on the interior Riemann sheet should guide the 
rays around the branch points 
and should define close loops or trajectories. Therefore, according to 
classical dynamics \cite{landau}, the number 
of potential or index profiles was strongly reduced to a few ones: 
harmonic-oscillator, Kepler profile and Maxwell's fish eye. In contrast, 
our proposed design is flexible enough to potentially generate a large 
variety of bounded orbits for light rays using 
striking properties of negative refractive indices. 

More importantly, our device based on optically isotropic material with negatively refracting 
index represents an example where the time delay is zero. Furthermore, due to impedance 
matching of negatively refracting materials, the waves are completely transmitted 
when crossing different material index borders, and reflecting waves are close to zero. These results 
strongly indicate that perfect invisibility with isotropic materials is possible.

We believe that our proposed design may stimulate theoretical and practical 
studies for designing and 
manufacturing cloaking devices using the properties of
negative refraction index. Emerging research based on metamaterials has 
opened up new avenues by exploiting the
artificial dielectric media. From telecommunications, radar 
and optical invisibility in defense areas to medical imaging and 
microelectronics in industrial sectors, 
the new emergent electromagnetic features obtained using metamaterials can 
bring exciting scientific progress. Further progress toward three-dimensional 
variations of this scheme is encouraged together with 
experimental work based on cloaking devices for achieving invisibility 
in the visible range of the spectrum.

\end{document}